\documentclass[runningheads]{llncs}
\usepackage{graphicx}
\usepackage{hyperref}
\usepackage{amsmath,amssymb}
\begin{document}
\title{Evidential segmentation of 3D PET/CT  images\thanks{This work was supported by the China Scholarship Council (grant 201808331005). It was carried out in the framework of the Labex MS2T  (Reference ANR-11-IDEX-0004-02)}}

\titlerunning{Evidential segmentation of 3D PET/CT images}

%
\author{Ling Huang\inst{1,2} \and
Su Ruan\inst{2} \and
Pierre Decazes\inst{3} \and
Thierry Den{\oe}ux\inst{1,4}
}
\authorrunning{L. Huang et al.}
%
\institute{Universit\'e de technologie de Compi\`egne, CNRS, Heudiasyc, Compi\`egne, France \\
\email{ling.huang@utc.fr} \and
University of Rouen Normandy, Quantif, LITIS, Rouen, France \and
CHB Hospital, Rouen, France \and
Institut universitaire de France, Paris, France }
\maketitle              
\begin{abstract}
PET and CT are  two modalities widely used in medical image analysis. Accurately detecting and segmenting lymphomas from these two imaging modalities are critical tasks for cancer staging and radiotherapy planning. However, this task is still challenging due to the complexity of PET/CT images, and the computation cost to process 3D data. In this paper, a segmentation method based on belief functions is proposed to segment lymphomas in 3D PET/CT images. The architecture is composed of a feature extraction module and an evidential segmentation (ES) module. The ES module outputs not only segmentation results (binary maps indicating the presence or absence of lymphoma in each voxel) but also uncertainty maps  quantifying the classification uncertainty. The whole model is optimized by minimizing Dice and uncertainty loss functions to increase segmentation accuracy. The method was evaluated on a database of 173 patients with diffuse large b-cell lymphoma. Quantitative and qualitative results show that our method outperforms the state-of-the-art methods.

\keywords{ lymphoma segmentation \and 3D PET/CT \and belief functions \and Dempster-Shafer theory \and uncertainty quantification \and deep learning}
\end{abstract}
\section{Introduction}

Positron Emission Tomography - Computed Tomography (PET/CT) scanning is an effective imaging tool for lymphoma segmentation with application to clinical diagnosis and radiotherapy planning. The standardized uptake value (SUV) for PET images is widely used to locate and segment lymphomas thanks to its high sensitivity and specificity to the metabolic activity of tumor. CT images are usually used jointly with PET images because of their anatomical feature representation capability.

\begin{figure}
\includegraphics[width=\textwidth]{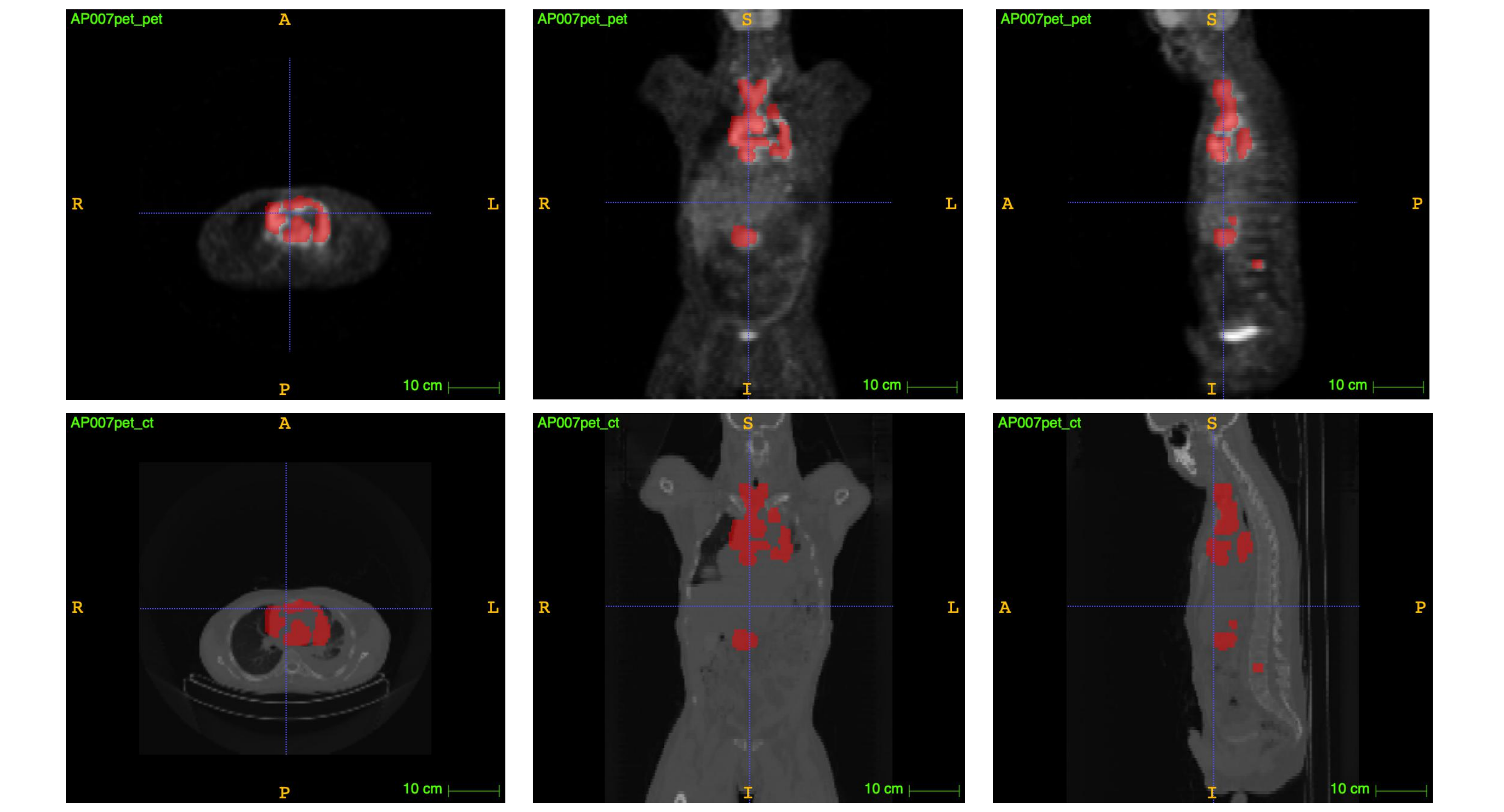}
\caption{Examples of patient with lymphomas. The first and second rows show, respectively PET and CT, slices of one patient in axial, sagittal and coronal views. The lymphomas are marked in red.}
\label{fig1}
\end{figure}

Although a lot of progress has been made in computer-aided lymphoma segmentation, the segmentation of whole-body lymphomas  is still challenging. (Fig.~\ref{fig1} shows an example of lymphoma patient. There is great variation in intensity distribution, shape, type and number of lymphomas). The methods can be classified into three main categories:  SUV-threshold-based \cite{ilyas2018defining},  region-growing-based \cite{hu2019detection} and  Convolutional Neural Network (CNN)-based \cite{li2019densex} methods. For PET images, it is common to segment lymphomas with a set of fixed SUV thresholds. This method is fast but lacks of flexibility in boundary delineation and requires domain knowledge to locate the region of interest. Region-growing-based methods have been proposed to optimize boundary delineation by taking texture and shape information into account. However, those methods still need clinicians to locate the seeds for region growing \cite{onoma2014segmentation}.

CNN-based segmentation methods have recently achieved great success. The UNet architecture \cite{ronnebergerconvolutional} has become the most popular medical image segmentation model. Driven by different tasks and datasets, many extended and optimized variants of UNet have been proposed, such as VNet \cite{milletari2016v}, nnUNet \cite{isensee2018nnu} and SegResNet \cite{myronenko20183d}. In \cite{li2019densex}, Li et al. propose a SegResNet-based lymphoma segmentation model with a two-flow architecture (segmentation and reconstruction flows). In \cite{blanc2020fully}, Blanc-Durand et al. propose a nnUNet-based lymphoma segmentation network.

Because of low resolution and contrast due to limitations of medical imaging technology,  PET/CT image segmentation results are tainted with uncertainty. 
Belief function (BF) theory \cite{shafer1976mathematical}\cite{denoeux20b}, also known as Dempster-Shafer theory, is a formal theory for information modeling, evidence combination and decision-making under uncertainty. In this paper, we propose a 3D PET/CT diffuse large b—cell lymphoma segmentation model based on BF theory and deep learning. 
The proposed deep neural network architecture is composed of a UNet module for feature extraction and a BF module for decision with uncertainty quantification. End-to-end learning is achieved by minimizing a two-part loss function allowing us to increase the Dice score while decrease the uncertainty. The model will first be described in Section \ref{sec:model} and experimental results will be reported in Section \ref{sec:results}.

\section{Methods}
\label{sec:model}

\subsection{Network Architecture}

\begin{figure}
\includegraphics[width=\textwidth]{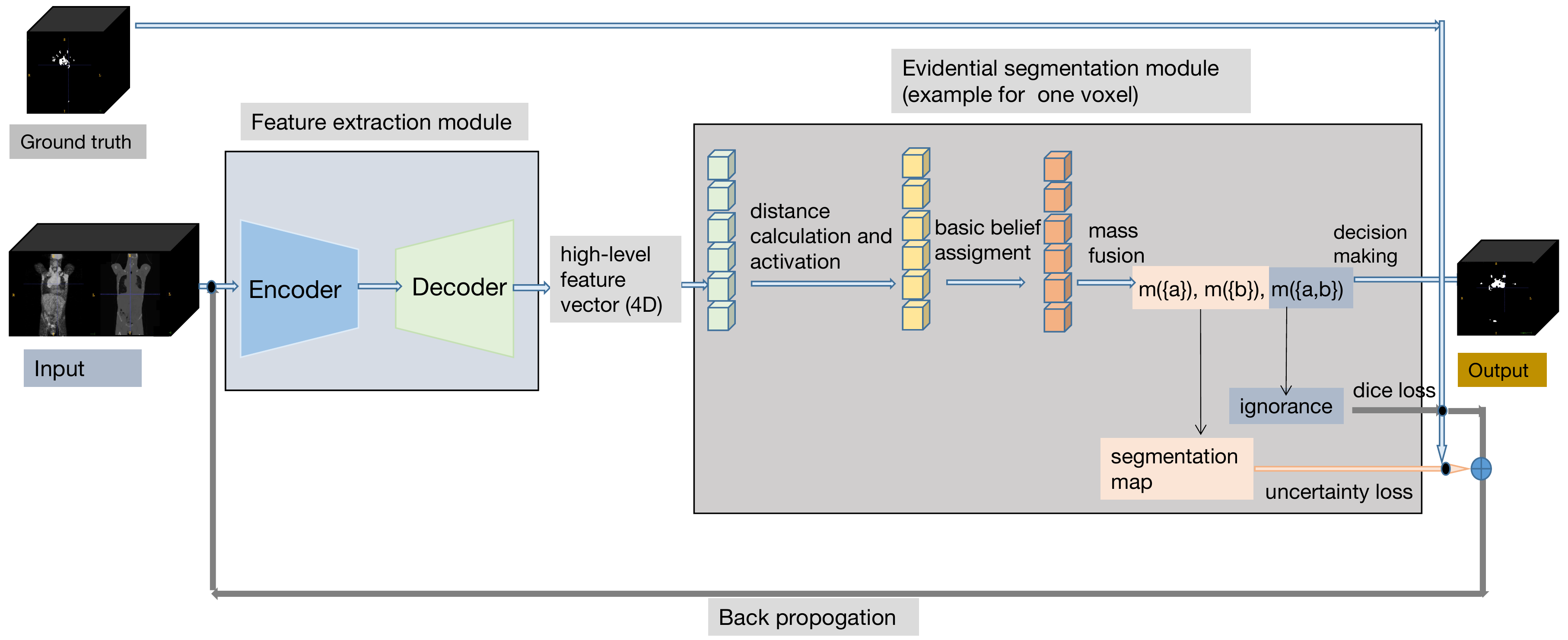}
\caption{Global lymphoma segmentation model (ES-UNet).}
\label{fig2}
\end{figure}

Fig.~\ref{fig2} shows the global lymphoma segmentation architecture (ES-UNet). It is composed of (1) an encoder-decoder feature extraction module (UNet), and (2) an evidential segmentation (ES) module comprising a distance activation layer, a basic belief assignment layer and a mass fusion layer. Details about the ES module will be given in Section \ref{subsec:ES}. Two loss terms are used for optimizing the training process: the \emph{Dice loss}, which quantifies the segmentation accuracy and  the \emph{uncertainty loss}, which quantifies the segmentation uncertainty. These loss functions will be described in Section \ref{subsec:loss}. A ``slim UNet'' with  $(8, 16, 32, 64, 128)$ convolution filters was implemented to reduce computation cost and  avoid overfitting.

\subsection{Evidential segmentation module}
\label{subsec:ES}

A probabilistic network with a softmax output layer may assign voxels a high probability of belonging to one class while the segmentation uncertainty is actually very high because, e.g., the voxel is located  close to the  fuzzy boundary between the tumor region and other tissues. Based on the evidential neural network model introduced in  \cite{denoeux2000neural} and using an  approach similar to that recently described in \cite{tong21a}, we propose a BF theory-based ES module to quantify the uncertainty about the class of each voxel by a Dempster-Shafer mass function. The main idea of the ES module is to assign a mass to each of the $K$ classes and to the whole set of classes $\Omega$, based on  the distance between the feature vector of each voxel and $I$ prototype centers. For a given voxel $x$, each prototype $p_{i}$ is considered as a piece of evidence, the reliability of which decreases with the Euclidean distance $d_{i}$ between $x$ and $p_{i}$. Each prototype $p_{i}$ is assumed to have a membership degree $u_{ik}$ to each class $\omega_k$ with the constraint $\sum_{k=1}^{K}u_{ik}=1$. The mass function induced by prototype $p_{i}$ is
\begin{subequations}
\label{eq:1}
\begin{align}
m_i(\{\omega_{k}\})&=\alpha _{i}u _{ik}\exp(-\gamma _{i}d_{i}^{2}), \quad k=1,\ldots,K\\
m_{i}(\Omega )&=1-\alpha _{i}\exp(-\gamma _{i}d_{i}^{2}), 
\end{align}
\end{subequations}
 The mass functions induced by the $I$ prototypes are then combined by Dempster's rule \cite{shafer1976mathematical}
\begin{equation}
m=\bigoplus _{i=1}^{I}m_{i}.
\label{eq:3}
\end{equation}

In our case, $\Omega=\{a, b\}$ and $K=2$. The ES module outputs for each voxel three mass values: two masses corresponding to lymphoma ($\{a\}$) and background ($\{b\}$), and an additional mass corresponding to  ignorance  ($\Omega$). For the voxels that are easy to classify into lymphoma or background, the mass values $m(\{a\})$ or $m(\{b\})$ are high and the mass $m(\{a, b\})$ is low. A high mass $m(\{a, b\})$ signals a lack of information to make a reliable decision.

\subsection{Loss function based on accuracy and uncertainty for segmentation}
\label{subsec:loss}

In general, a good segmentation system is expected to make few segmentation errors while providing as informative outputs as possible. Since we  quantify  uncertainty by the ``ignorance class'' via the evidential network, we propose to minimize a loss function defined as the sum of two terms: a Dice loss $\textsf{loss}_{d}$ that measures the discrepancy between the ground truth and segmentation outputs, and an uncertainty loss $\textsf{loss}_u$ that measures the uncertainty of the segmentation outputs. More precisely, the Dice loss is defined as
\begin{equation}
    \textsf{loss}_{d}=1-\frac{2 \sum_{n=1}^{N} S_n G_n}{ \sum_{n=1}^{N} S_n+ \sum_{n=1}^{N} G_n},
    \label{eq:4}
\end{equation}
where $N$ is the number of voxels in the image volume, $S$ is the segmentation outputs of our model and $G$ is the ground truth. The uncertainty loss is defined as
    \begin{equation}
        \textsf{loss}_{u}={\frac {1}{N}}\sum _{n=1}^{N} [m_n(\Omega)]^2,
    \label{eq:5}
    \end{equation}
where $m_n$ is the mass function computed for voxel $n$. The total loss function is then
    \begin{equation}
        \textsf{loss}=\textsf{loss}_{d}+\textsf{loss}_{u} +\lambda \left \| \alpha \right \|_1,  
    \label{eq:6}
    \end{equation}
where $\lambda$ is the regularization coefficient for  parameter vector $\alpha=(\alpha_1,\ldots,\alpha_I)$ with $\alpha_i$ defined in \eqref{eq:1}. The regularization term allows us to decrease the influence of unimportant prototype centers and avoid overfitting.


\section{Experimental results}
\label{sec:results}

\subsection{Experimental settings}
The dataset contains images from 173 patients who were diagnosed with large b-cell lymphoma and underwent PET/CT examination. The study was approved as a retrospective study by the Henri Becquerel Center Institutional Review Board. The lymphomas in mask images were delineated manually by experts and considered as ground truth. The size and spatial resolution of PET and CT images and the corresponding mask images vary due to different imaging machines and operations, which makes it difficult to transfer the data into a deep neural model directly. For prepossessing, we resized PET, CT and mask images to the same size $256\times256 \times128$. For normalization,
we set (shift, scale) values to (0, 0.1) and (1000, 1/2000) for, respectively  PET and CT images. 

We randomly selected 80\% of the data for training, 10\% for validation and 10\% for testing. Dice score, sensitivity, specificity, precision and F1 score were used to evaluate the segmentation performance. We first computed the five indices for each test patient and then averaged these indices over the patients. During training, PET and CT images were concatenated as a two-channel input. The prototype vectors and membership degrees where initialized randomly with uniform distributions, and the vectors \textsf{$\alpha$} and \textsf{$\gamma$} where initialized with constants 0.5 and 0.01. The number of prototypes was set to 20. The learning rate was set to $10^{-3}$ during training and the model was trained with 50 epochs using the Adam optimization algorithm. The regularization coefficient $\lambda$ in \eqref{eq:6} was set to $10^{-5}$.  All methods were implemented in Python with a PyTorch-based, medical image framework MONAI and were trained and tested on a desktop with a 2.20GHz Intel(R) Xeon(R) CPU E5-2698 v4 and a Tesla V100-SXM2 graphics card with 32 GB GPU memory. 

\subsection{Results and discussion}

\begin{figure}
\centering
\includegraphics[width=\textwidth]{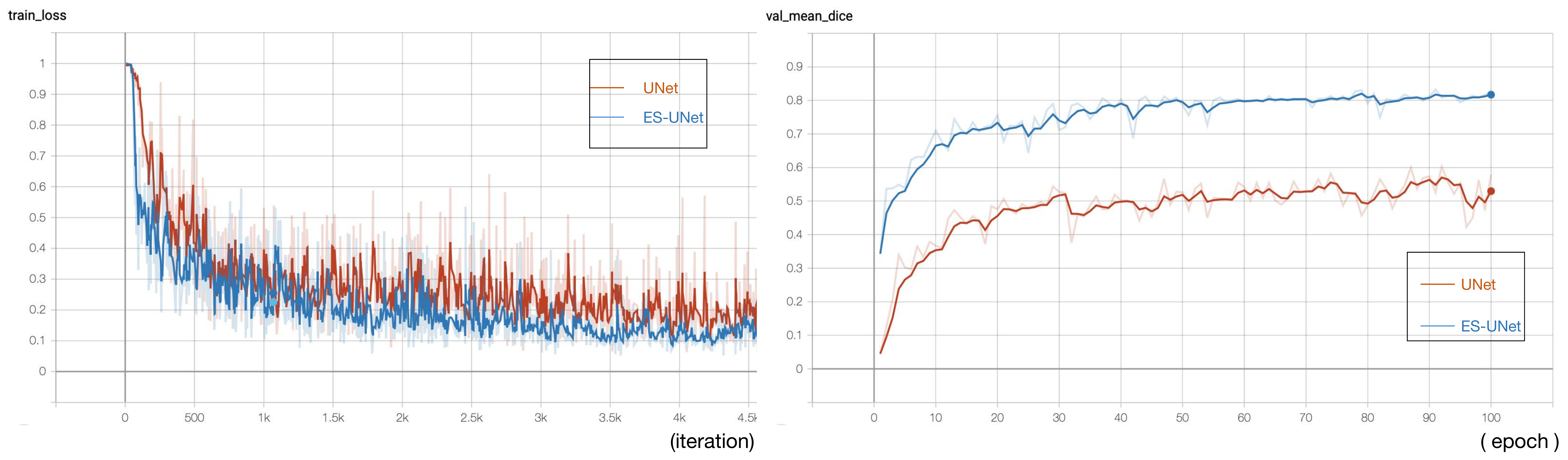}
\caption{Training process visualization: training loss (left) and validation Dice score (right).}
\label{fig3}
\end{figure}

\begin{figure}
\centering
\includegraphics[width=\textwidth]{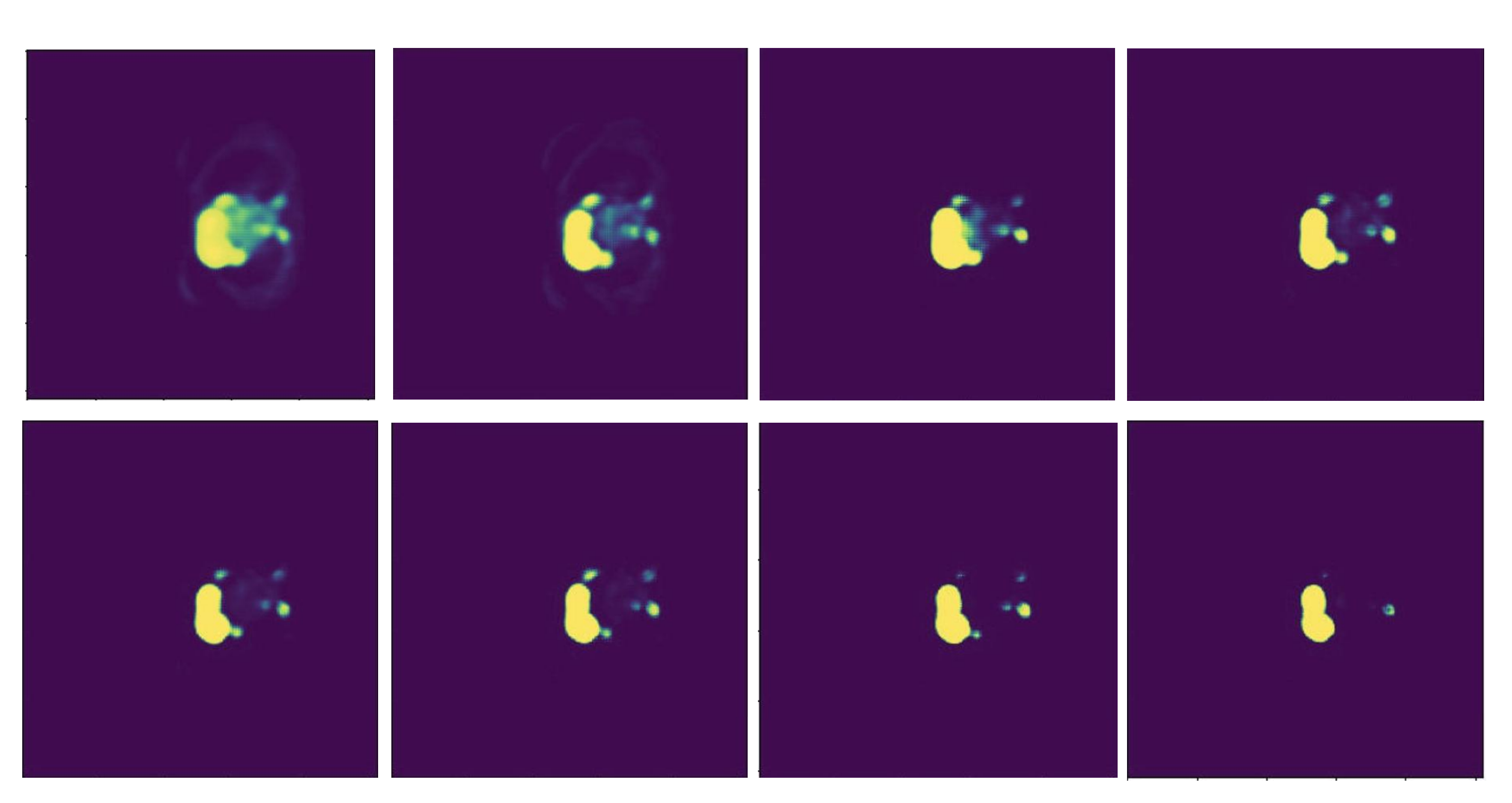}
\caption{Uncertainty maps obtained during the training. For one map, the pixels classified to background, lymphoma, ignorance are marked in purple, yellow and iridescent, respectively.}
\label{fig4}
\end{figure}

The quantitative results are shown in Table~\ref{tab1}. Our model outperforms the baseline model UNet as well as the other state-of-the-art methods. In particular, our model outperforms the best model SegResNet by, respectively, 1.9\%, 2.4\%, 1.4\% in Dice score, Sensitivity and F1 score. It should be noted that the state-of-the-art models were trained with 100 epochs on our dataset because they are slower to converge during training. Fig.~\ref{fig3} displays the learning curves of the training loss and validation Dice score for UNet and ES-UNet, showing the advantage of ES-UNet in terms not only of segmentation accuracy, but also of convergence speed. Fig.~\ref{fig4} shows the segmentation and uncertainty maps at different steps during the training of ES-UNet. Our model quantifies the uncertainty of ambiguous pixels instead of classifying them unambiguously into a single class. The uncertainty decreases during the learning process thanks to the minimization of the uncertainty loss term.

\begin{table}
\caption{Performance comparison with the baseline methods on the test set.}
\centering
\label{tab1}
\begin{tabular}{|l|l|l|l|l|l|}
\hline
Models  &Dice score &Sensitivity&  Specificity & Precision & F1 score\\
\hline
ES-UNet (our model) & \textbf{0.830} & 0.923 & 0.908 &0.912 & \textbf{0.915}\\
UNet \cite{ronnebergerconvolutional} & 0.769& 0.798 & \textbf{0.963} &0.890&0.833\\
nnUNet \cite{isensee2018nnu} & 0.702 & \textbf{0.950} & 0.499 &0.758 & 0.807\\
VNet \cite{milletari2016v}&0.802&0.882 & 0.904 & 0.916&0.909\\
SegResNet \cite{myronenko20183d}& 0.811 & 0.899 & 0.942 & \textbf{0.925} & 0.901\\
\hline
\end{tabular}
\end{table}

\begin{figure}
\centering
\includegraphics[width=\textwidth]{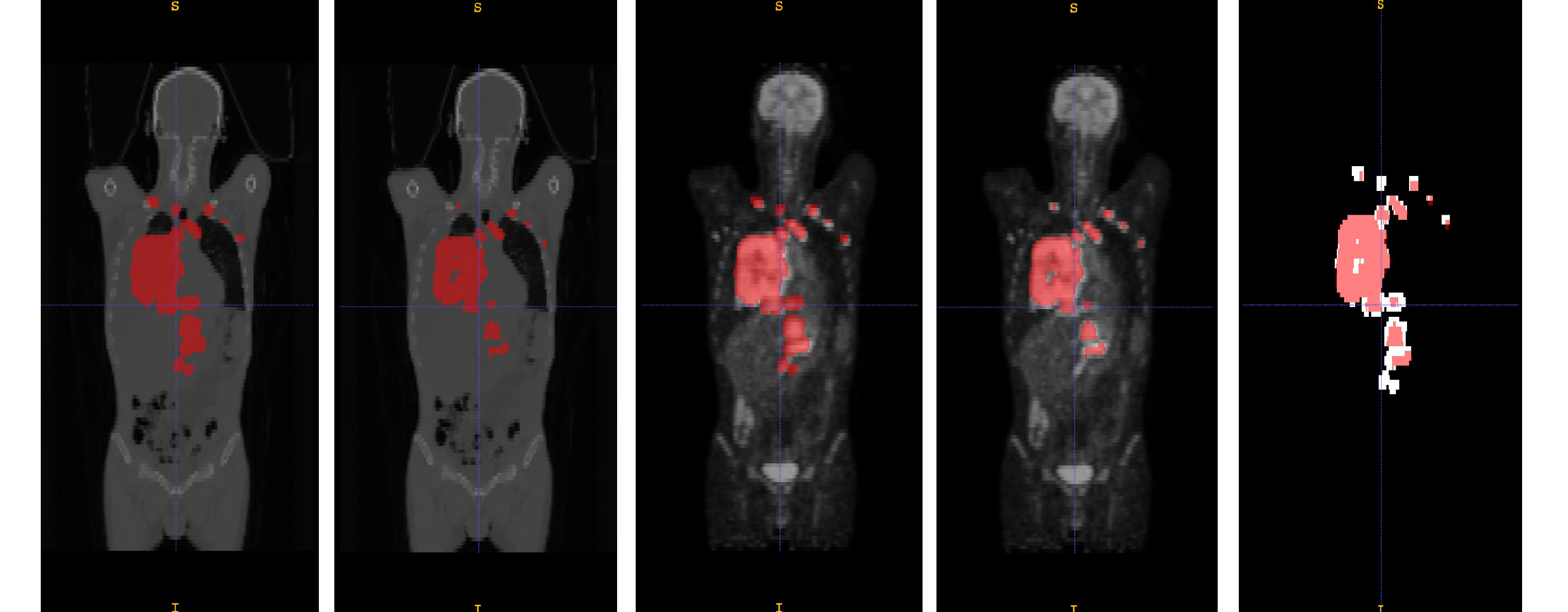}
\caption{Segmentation results of ES-UNet. From left to right: ground truth, segmentation results, difference map between the ground truth, and segmented lymphomas.}
\label{fig5}
\end{figure}

Fig.~\ref{fig5} shows an example of  segmentation results obtained by ES-UNet. Our model can locate and segment most of the lymphomas. The segmentation results were found credible and were confirmed by experts.

\section{Conclusion}
An evidential segmentation framework (ES-UNet) for segmentation of lymphomas from 3D PET/CT with uncertainty quantification has been introduced. The proposed architecture is based on the concatenation of a UNet and an evidential segmentation layer, making it possible to compute output mass functions for each voxel. The training is performed by minimizing a two-part loss function composed of a Dice loss and an uncertainty loss, with the effect of increasing the Dice score while decreasing the uncertainty. Qualitative and quantitative evaluations show promising results when compared to the baseline model UNet as well as the state-of-the-art methods. Future research will tackle multi-modality medical image fusion with BF theory. 

%


%
%

%
%
%
 \bibliographystyle{splncs04}
 \bibliography{ref}
%




\end{document}